\documentclass{rnaastex}

\begin{document}

\title{Kozai-Lidov Resonant Behavior among Atira-class Asteroids}

\correspondingauthor{Carlos~de~la~Fuente~Marcos}
\email{nbplanet@ucm.es}

\author[0000-0003-3894-8609]{Carlos~de~la~Fuente~Marcos}
\affiliation{Universidad Complutense de Madrid \\
             Ciudad Universitaria, E-28040 Madrid, Spain}

\author[0000-0002-5319-5716]{Ra\'ul~de~la~Fuente~Marcos}
\affiliation{AEGORA Research Group \\
             Facultad de Ciencias Matem\'aticas \\
             Universidad Complutense de Madrid \\
             Ciudad Universitaria, E-28040 Madrid, Spain}


\section{} 

Atira-class asteroids have aphelia greater than 0.718~au and smaller than 0.983~au \citep{2012Icar..217..355G}; as of 29-May-2018, there are
only 18 known Atiras because, like Venus, they can only be properly observed near their greatest elongation. \citet{2016MNRAS.458.4471R} 
have found that the orbits of many of them remain relatively stable for at least 1~Myr, but it is unclear what can make them so stable when 
they cross the orbit of Venus and some may reach, in heliocentric terms, as far as Earth's perihelion and as close as inside Mercury's 
perihelion. The analysis of their orbits shows that many of them have values of the argument of perihelion, $\omega$, close to 0{\degr} or 
180{\degr}. Such an orbital arrangement means that the nodes are located at perihelion and at aphelion \citep{1989Icar...78..212M}, i.e. for
many known Atiras, away from the path of Venus. Another remarkable trend is in the eccentricity--inclination ($e-i$) relationship; Atiras 
with $e<0.34$ have $i>10\degr$ while those with $e>0.34$ tend to have lower values of $i$. Although any clustering in $\omega$ could be due 
to observational bias, the trend in $e-i$ may be signaling a Kozai-Lidov scenario in which both the values of $e$ and $i$ oscillate, 
alternating high $e$ and $i$, because the value of the so-called Kozai-Lidov parameter, $\sqrt{1-e^{2}}\ \cos{i}$ remains constant 
\citep{1962AJ.....67..591K,1962P&SS....9..719L}. For regular heliocentric ecliptic orbital elements, the Kozai-Lidov parameter measures the 
evolution of the component of the orbital angular momentum of the object perpendicular to the ecliptic. Within the Kozai-Lidov resonance, 
many known Atiras would be protected against close encounters with Mercury, Venus, and the Earth--Moon system. Here, we show that the orbits 
of some Atiras are strongly affected by the Kozai-Lidov resonance.

We focus on the evolution of (163693)~Atira ($\omega=253\degr$), (413563) 2005~TG$_{45}$ ($\omega=230\degr$) and (418265) 2008~EA$_{32}$ 
($\omega=182\degr$) ---three well-observed Atiras--- and also on 2010~XB$_{11}$ ($\omega=202\degr$), 2013~TQ$_{5}$ ($\omega=247\degr$), and 
the most recent addition to the Atira-class, 2018~JB$_{3}$ ($\omega=355\degr$). Our results are based on the latest orbital solutions 
available from \href{http://ssd.jpl.nasa.gov/sbdb.cgi}{JPL's Small-Body Database}. Our full $N$-body simulations have been performed as 
described by \citet{2012MNRAS.427..728D}. 

\begin{figure}[!ht]
\begin{center}
\includegraphics[scale=0.55,angle=0]{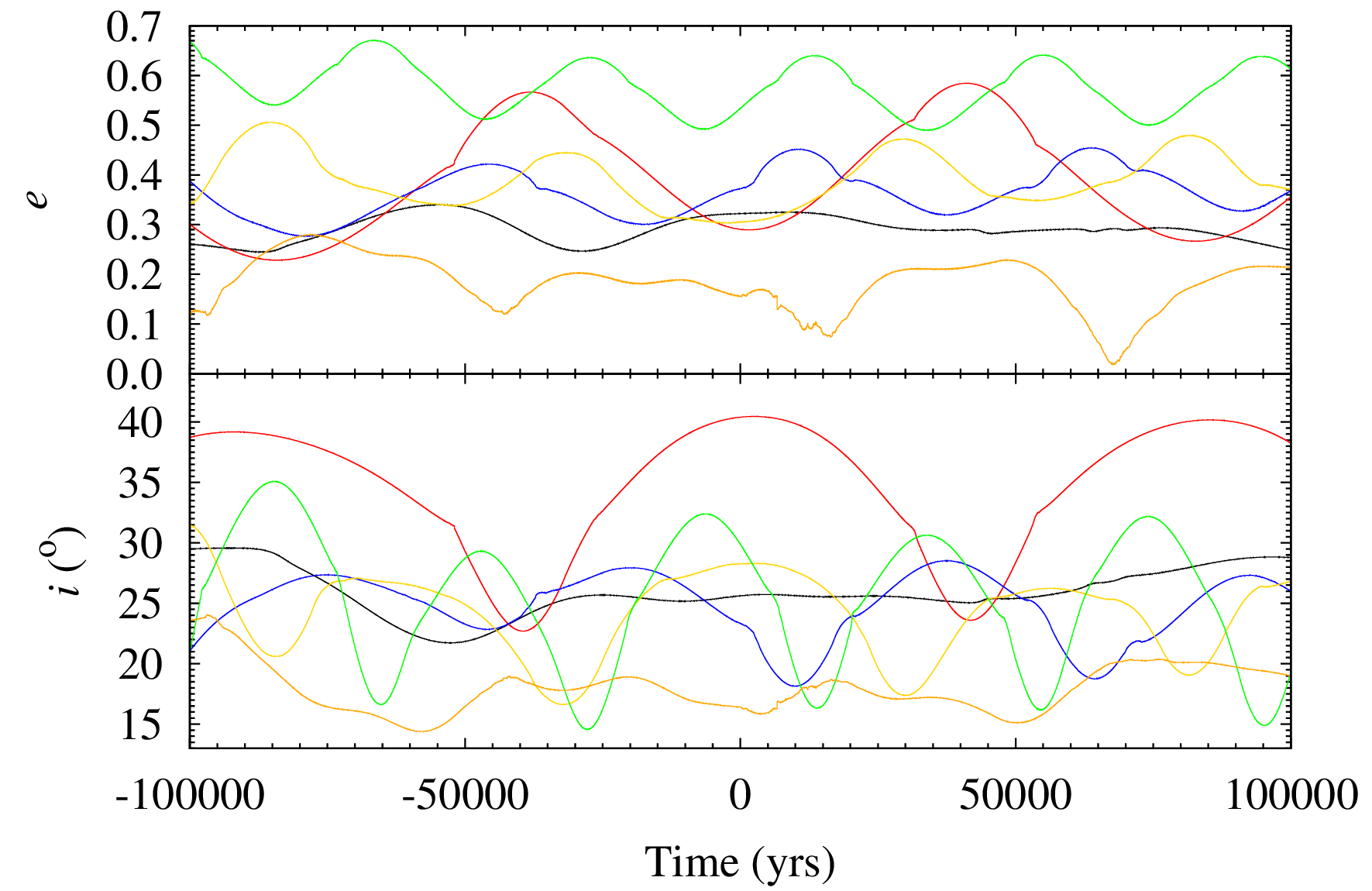}
\includegraphics[scale=0.55,angle=0]{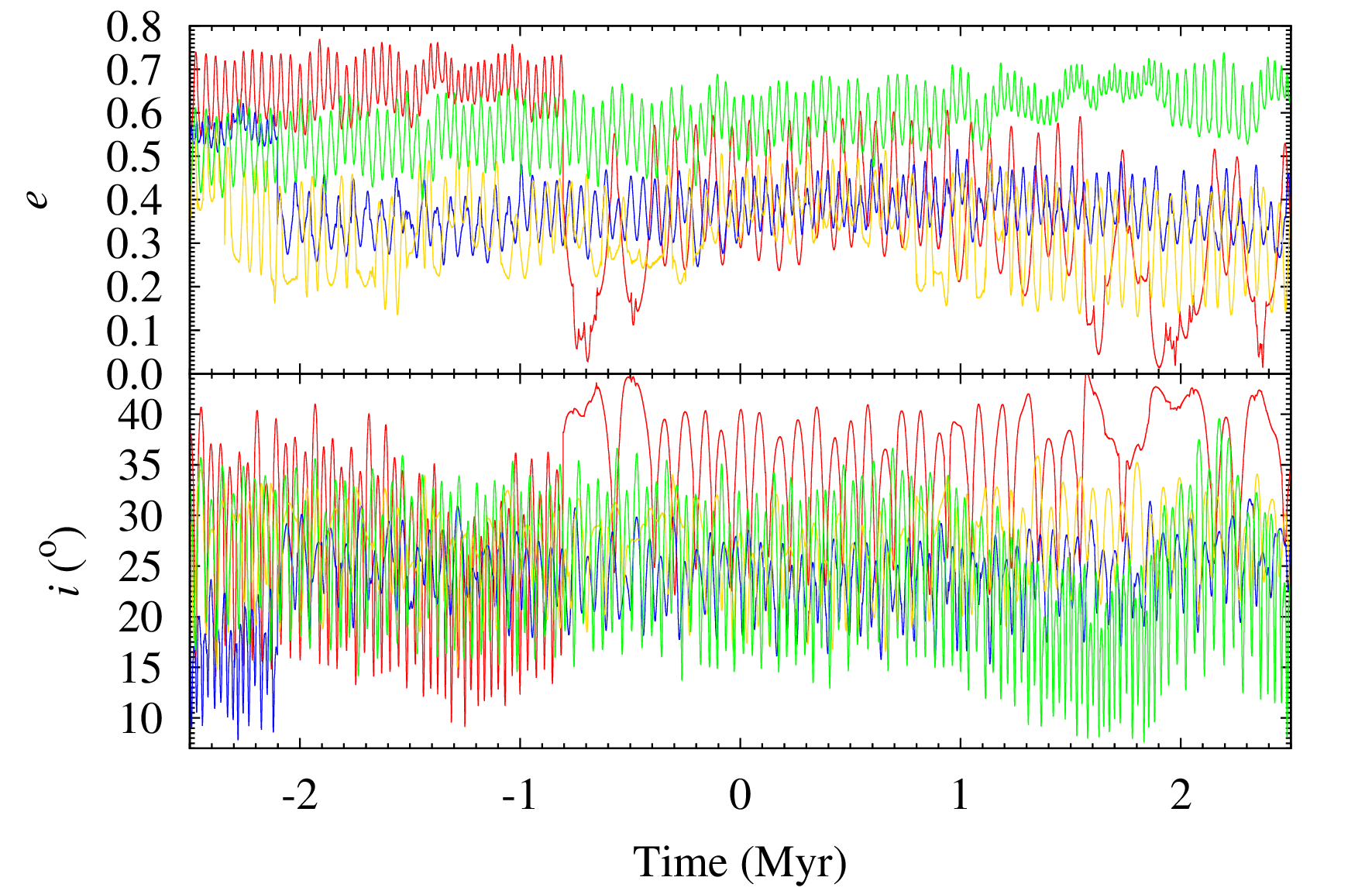}
\caption{Short-term evolution of the values of the eccentricity (top panel) and inclination (second to top panel) of the nominal orbits 
         (zero instant of time, epoch JDTDB~2458200.5, 23-March-2018) of (163693)~Atira (black), (413563) 2005~TG$_{45}$ (blue), (418265) 
         2008~EA$_{32}$ (gold), 2010~XB$_{11}$ (green), 2013~TQ$_{5}$ (orange), and 2018~JB$_{3}$ (red). For 413563, 418265, 2010~XB$_{11}$ 
         and 2018~JB$_{3}$, eccentricity and inclination exhibit the interplay characteristic of the Kozai-Lidov resonance. The two
         additional bottom panels show a longer time-span. 
\label{fig:1}}
\end{center}
\end{figure}

Figure~\ref{fig:1} shows the short-term evolution of $e$ (top panel) and $i$ (second to top panel). Atira and 2013~TQ$_{5}$ exhibit a rather 
irregular behavior; in sharp contrast, 413563, 418265, 2010~XB$_{11}$, and 2018~JB$_{3}$ show the coupling in $e-i$ characteristic of the 
Kozai-Lidov mechanism. Such a behavior is often found among near-Earth asteroids (NEAs) with values of their semi-major axes close to that 
of Earth that move confined between the orbits of Venus and Mars \citep{1996A&A...307..310M,2015A&A...580A.109D}, but in this case the value 
of $\omega$ also oscillates. We do not find such a libration in $\omega$ for the Atiras (not shown). It can hardly be argued that the 
population of NEAs trapped in the Kozai-Lidov resonance is a good dynamical analogue of many Atiras, but the lack of libration in $\omega$ 
might be used as an argument against such an analogy. \citet{2009A&A...493..677L} have shown that within the context of the Kozai-Lidov 
resonance, $\omega$ may circulate in the general reference frame, but it always librates in the Laplace-plane reference frame. Our 
calculations are referred to the usual ecliptic and mean equinox of reference epoch, not to the invariable plane of the Solar System and 
this is why the oscillation in $\omega$ is not observed. Therefore, what is found is consistent with a Kozai-Lidov resonant state.

In this Note, we have explored the orbital evolution of representative Atira-class asteroids in an attempt to understand better the results
discussed in \citet{2016MNRAS.458.4471R}. Our $N$-body calculations and subsequent analysis show that many known Atiras undergo 
Kozai-Lidov-type orbital evolution, which translates into enhanced long-term stability (see Figure~\ref{fig:1}, bottom two panels). 

\acknowledgments

We thank S.~J. Aarseth for providing the code used in this research, F. Roig, M.~N. De Pr{\'a}, and S. Deen for comments, and A.~I. G\'omez 
de Castro for providing access to computing facilities. This work was partially supported by the Spanish MINECO under grant ESP2015-68908-R. 
In preparation of this Note, we made use of the NASA Astrophysics Data System and the MPC data server.


\begin{thebibliography}{}

\bibitem[de la Fuente Marcos \& de la Fuente Marcos(2012)]{2012MNRAS.427..728D} de la Fuente Marcos, C., \& de la Fuente Marcos, R.\ 2012, \mnras, 427, 728

\bibitem[de la Fuente Marcos \& de la Fuente Marcos(2015)]{2015A&A...580A.109D} de la Fuente Marcos, C., \& de la Fuente Marcos, R.\ 2015, \aap, 580, A109

\bibitem[Greenstreet et al.(2012)]{2012Icar..217..355G} Greenstreet, S., Ngo, H., \& Gladman, B.\ 2012, \icarus, 217, 355

\bibitem[Kozai(1962)]{1962AJ.....67..591K} Kozai, Y.\ 1962, \aj, 67, 591

\bibitem[Libert \& Tsiganis(2009)]{2009A&A...493..677L} Libert, A.-S., \& Tsiganis, K.\ 2009, \aap, 493, 677

\bibitem[Lidov(1962)]{1962P&SS....9..719L} Lidov, M.~L.\ 1962, \planss, 9, 719

\bibitem[Michel \& Thomas(1996)]{1996A&A...307..310M} Michel, P., \& Thomas, F.\ 1996, \aap, 307, 310

\bibitem[Milani et al.(1989)]{1989Icar...78..212M} Milani, A., Carpino, M., Hahn, G., \& Nobili, A.~M.\ 1989, \icarus, 78, 212

\bibitem[Ribeiro et al.(2016)]{2016MNRAS.458.4471R} Ribeiro, A.~O., Roig, F., De Pr{\'a}, M.~N., Carvano, J.~M., \& DeSouza, S.~R.\ 2016, \mnras, 458, 4471

\end{thebibliography}
\end{document}